\documentclass[prl,twocolumn,superscriptaddress,amsmath,amssymb,showpacs,floatfix,preprintnumbers]{revtex4-2}

\usepackage{amsmath}
\usepackage{graphicx}
\usepackage{dcolumn}
\usepackage{xcolor}
\usepackage{physics}
\usepackage{xr}
\usepackage{siunitx}
\usepackage{tabularx}
\usepackage{makecell}
\usepackage{multirow}
\DeclareGraphicsExtensions{.png .jpg .pdf}

\usepackage[normalem]{ulem}

\makeatletter
\newcommand*{\addFileDependency}[1]{
  \typeout{(#1)}
  \@addtofilelist{#1}
  \IfFileExists{#1}{}{\typeout{No file #1.}}
}
\makeatother
 
\newcommand*{\myexternaldocument}[1]{%
    \externaldocument{#1}%
    \addFileDependency{#1.tex}%
    \addFileDependency{#1.aux}%
}

\myexternaldocument{si}

\begin{document}

\title{Moir\'e Kramers-Weyl Fermions from Structural Chirality with Ideal Radial Spin Texture}

\author{D. J. P. de Sousa}\thanks{These authors contributed equally to this work.}
\author{Seungjun Lee}\thanks{These authors contributed equally to this work.}
\affiliation{Department of Electrical and Computer Engineering, University of Minnesota, Minneapolis, Minnesota 55455, USA}

\author{Tony Low}\email{tlow@umn.edu}
\affiliation{Department of Electrical and Computer Engineering, University of Minnesota, Minneapolis, Minnesota 55455, USA}
\affiliation{Department of Physics, University of Minnesota, Minneapolis, Minnesota 55455, USA}

\begin{abstract}
We demonstrate that two-dimensional Kramers-Weyl fermions can be engineered in spin-orbit coupled twisted bilayers, where the chiral structure of these moir\'e systems breaks all mirror symmetries, confining Kramers-Weyl fermions to high-symmetry points in the Brillouin zone under time reversal symmetry. Our theoretical analysis reveals a symmetry-enforced Weyl-like interlayer moiré coupling that universally ensures an ideal radial spin-texture at arbitrary twist angles, under $C_{nz}$ symmetry with $n>2$. First principles density functional calculation confirm the realization of these fermions in twisted $\alpha$-In$_2$Se$_3$ bilayers, where flat bands and out-of-plane ferroelectric polarization in each layer guarantee two-dimensional Kramers-Weyl physics with perfectly ideal radial spin textures.
\end{abstract}

\maketitle

\emph{Introduction.} In three-dimensional non-magnetic chiral crystals, the presence of strong spin-orbit interaction inherently gives rise to electronic states hosting topological Kramers-Weyl fermions~\cite{Chang2018, Bradlyn2016, PhysRevLett.130.066402, Rao2019, Schrter2019, PhysRevLett.119.206401, PhysRevLett.119.206402, Ma2021, PhysRevB.108.L201404, PshenaySeverin2018, Yao2020, Shekhar2018,PhysRevLett.131.116603, Yang2023, He2021, PhysRevB.108.L201404, PhysRevResearch.3.033101}. These quasiparticles are endowed with an intrinsic chirality, whose handedness derives from the structural chirality of the real space crystal lattice, manifesting through a radial spin-texture~\cite{Chang2018, PhysRevLett.125.216402, Tan2022}. In two-dimensional (2D) non-magnetic systems, typically reported Weyl states such as in few layers 1T$^{\prime}$-W(Mo)Te$_2$~\cite{Li2017, Tang2017, Deng2016, PhysRevB.94.121113} or $\alpha$-Bi based van der Waals heterostructures~\cite{https://doi.org/10.48550/arxiv.2303.02971, https://doi.org/10.48550/arxiv.2311.18026}, are not Kramers-Weyl semimetals. The emergence of Weyl fermions in these systems arises primarily from the combination of strong spin-orbit coupling and broken inversion symmetry. The latter allows for the existence of mirror planes, which provide protection to band crossings away from time reversal invariant momenta (TRIM) in the Brillouin zone~\cite{PhysRevLett.119.226801}. Hence, true Kramers-Weyl fermions derived directly from intrinsically chiral structures in 2D systems, lacking all mirror symmetries, deserve further investigation. 

The field of twistronics has unveiled a novel paradigm for probing the impact of structural chirality on the electronic characteristics of two-dimensional quantum materials~\cite{PhysRevB.95.075420, 2024, Zhu2024, Andrei2020, Jorio2022, Oh2021, Cao2018,2Cao2018,https://doi.org/10.1002/adma.202310768}. The moir\'e electronic states intrinsically inherit the structural chirality of twisted atomic bilayers through the interlayer moiré coupling, thereby enabling unprecedented opportunities for manipulating quantum phenomena by exploiting the twist degree of freedom~\cite{PhysRevB.106.165420, PhysRevB.106.L081406, https://doi.org/10.48550/arxiv.2312.10227, Kim2016, PhysRevLett.125.077401}. While the influence of structural chirality on moir\'e electron states has garnered intense scientific interest, its impact on the electronic properties of strongly spin-orbit coupled systems warrant further investigation. Consequently, the prospects of realizing genuine Kramers-Weyl semimetals in 2D structurally chiral structures presents exciting opportunities topological semimetallic states in moir\'e systems.

In this letter, we show that strongly spin-orbit coupled moir\'e states can host truly chiral Kramers-Weyl fermions in twisted atomic bilayers. Our findings are supported by a symmetry-based continuum model as well as first principles electronic structure calculations in a prototypical system; twisted $\alpha$-In$_2$Se$_3$ bilayers. The interplay between structural chirality, strong spin-orbit interaction and time-reversal symmetry in twisted $\alpha$-In$_2$Se$_3$ bilayers gives rise to moir\'e Kramers-Weyl states at TRIM in the Brillouin zone, exhibiting ideal radial spin textures \textit{at arbitrary twist angles}. The 2D radial spin texture of moir\'e Kramers-Weyl states is found to be strongly enhanced by virtue of the out-of-plane ferroelectric polarization within each $\alpha$-In$_2$Se$_3$. While each isolated $\alpha$-In$_2$Se$_3$ monolayer displays a Rashba spin-texture, our findings reveal that the structural chirality unavoidably leads to ideal Weyl-type interlayer moir\'e coupling for valence electrons. We also show that this result is generally true in the presence of $C_{3z}$, $C_{4z}$ and $C_{6z}$ symmetries, while interlayer moir\'e coupling displaying chiral radial-tangential spin texture~\cite{PhysRevB.104.104408} is possible otherwise. We discuss how the existence of ultra-flat valence bands in these systems~\cite{tao2022designing} offer the possibiltiy of realizing ideal moir\'e Kramers-Weyl fermions. This work elucidates the role of structural chirality, spin-orbit interaction and time-reversal symmetry in enabling emergent electron states in engineered moir\'e quantum materials.

\begin{figure*}[t]
\includegraphics[scale = 0.9]{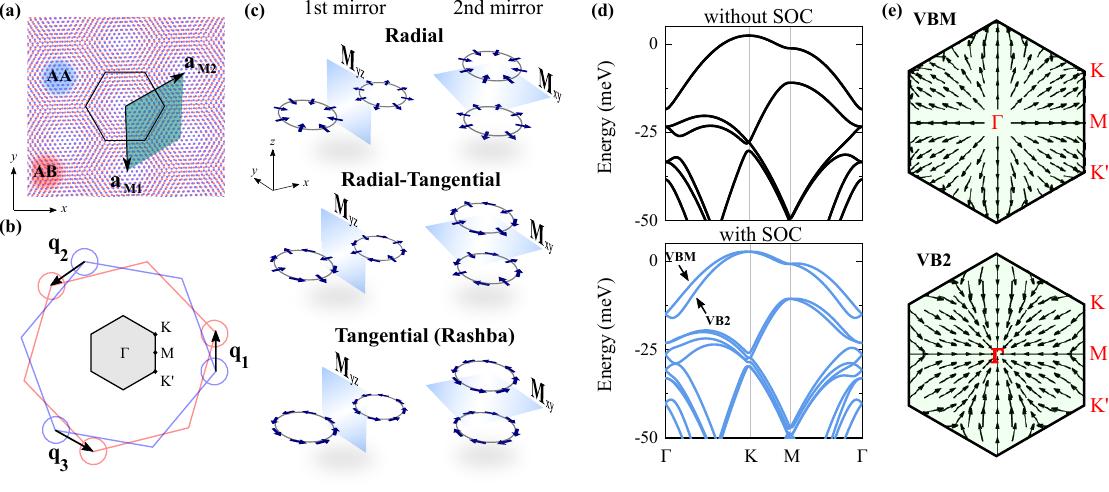}
\caption{(a) Moir\'e pattern of twisted lattices made from the Wannier $p_z$ orbitals. AA and AB local stacking regions as well as the moir\'e unit cell are highlighted. (b) Three smallest transfer momenta employed in the continuum model. The moir\'e Brillouin zone is shows at the center. (c) Transformation of radial (top), tangential-radial (middle) and Rashba (bottom) spin textures under vertical, $\mathcal{M}_{yz}$, and horizontal, $\mathcal{M}_{xy}$, mirror operations. (d) Electronic structure obtained from the continuum model without (top) and with (bottom) spin-orbit coupling. (e) Spin texture of the first two valence bands, highlighted in panel (e), over the whole moir\'e Brillouin zone.}
\label{fig2}
\end{figure*}

\emph{Moir\'e States in Strongly Spin-Orbit Coupled Systems.}  
We begin by discussing the moir\'e physics of strongly spin-orbit coupled twisted bilayers. Figure~\ref{fig2}(a) displays the moir\'e pattern made from twisted triangular lattices of localized $p_z$ Wannier states assumed here, without loss of generality. The moir\'e lattice vectors are $\textbf{a}_{M1} = (4\pi/\sqrt{3}k_{\theta})(0, -1)$ and $\textbf{a}_{M2} =  (4\pi/\sqrt{3}k_{\theta})(\sqrt{3}/2, 1/2)$ with $k_{\theta} = (8\pi/\sqrt{3}a_0)\sin(\theta/2)$ and $a_0$ is the lattice constant of each monolayer. The moir\'e unit cell as well as the inequivalent local stacking regions are highlighted in Fig.~\ref{fig2}(a). Next, we derive the moir\'e Hamiltonian for our model system.  For concreteness, we consider the specific symmetries of $\alpha$-In$_2$Se$_3$ systems. 

We start by deriving the intralayer Hamiltonians. We assume that the symmetries of the $p_z$ Wannier states within each monolayer are a subset of the space group $D_{6h}$ operations, containing two rotation operations $C_{2z}$, $C_{3z}$ and two mirror planes $\mathcal{M}_{xz}$, $\mathcal{M}_{yz}$~\cite{Snote}. We assume that the system is time-reversal symmetric and that inversion symmetry in recovered in the bilayer structure. The continuum Hamiltonian for top (T) and bottom (B) layers around the K points is, therefore, given by
\begin{eqnarray}
H^{\eta T(B)}(\textbf{k}) = \epsilon_0 s_0 + \lambda^{\eta T(B)}_{R}[k_x s_y - k_ys_x] + \gamma k^2 s_0,
    \label{eq1}
\end{eqnarray}
to second order in $\textbf{k}$, where $\gamma$ is a negative constant related to the effective mass of $p_z$ states at the valence band and $\eta = \pm 1$ is the valley index. The $\mathcal{M}_{yz}$ symmetry requires $\lambda_R^{\eta T(B)} = \lambda_R^{-\eta T(B)}$ and the inversion symmetry requires $\lambda^{\eta T}_R = -\lambda^{\eta B}_R$, resulting in a layer dependent Rashba spin texture with opposite chiralities in the two layers, but with same chirality at non-equivalent valleys within each layer. Next, we discuss the interlayer moir\'e coupling.

We follow Refs.~\cite{Bistritzer2011, PhysRevLett.123.036401} and write the interlayer moir\'e coupling as
\begin{eqnarray}
    \langle \Psi^{\eta}_{\alpha\textbf{k}} | \hat{H} | \Psi^{\eta}_{\alpha'\textbf{k}'} \rangle = \frac{1}{\Omega} \sum_{\textbf{G}\textbf{G}'} t^{\eta, \alpha \alpha'}_{\textbf{k} + \textbf{G}} e^{i \textbf{G} \cdot \boldsymbol{\tau}_{\alpha} } e^{-i \textbf{G}' \cdot \boldsymbol{\tau}'_{\alpha} } \delta_{\textbf{k} - \textbf{k}', \textbf{G}' - \textbf{G}}, \nonumber \\
    \label{eq11}
\end{eqnarray}
where $\Omega$ is the sample area and $\alpha, \textbf{k}$, $\textbf{G}$ ($\alpha', \textbf{k}'$, $\textbf{G}'$ ), refer to spin index, momentum and reciprocal lattice vectors, respectively, of the T (B) layer. The reciprocal lattice vectors characterize the interlayer momentum transfer $\textbf{q} = \textbf{G} - \textbf{G}'$. The most relevant momentum transfer are the three smallest ones: $\textbf{q}_1 = \textbf{K}^{T} - \textbf{K}^{B}$, $\textbf{q}_2 = C_{3z}\textbf{q}_1$ and $\textbf{q}_3 = C_{3z}\textbf{q}_2$, where $\textbf{K}^{T} = R(\theta)\textbf{K}$ and $\textbf{K}^{B} = \textbf{K}$ are the K points of T and B layers respectively (and similarly for the $K'$ point). Here, $R(\theta)$ is the rotation matrix and $\theta$ is the twist angle. The magnitudes of the interlayer momentum transfer are, therefore, given by $|\textbf{q}_j| = 2|\textbf{K}|\sin (\theta /2)$ for $j = 1,2,3$. The transfer momenta are depicted in Fig.~\ref{fig2}(b) along with the moir\'e Brillouin zone.

In moving forward, we write the interlayer moir\'e coupling in the compact form $ \langle \Psi^{\eta}_{\textbf{k}} | \hat{H} | \Psi^{\eta}_{\textbf{k}'} \rangle  = \sum_{\textbf{G}\textbf{G}'} W^{\eta}_{\textbf{G}\textbf{G}'}(\textbf{k})\delta_{\textbf{k} - \textbf{k}', \textbf{G}' - \textbf{G}}$ and explicitly write the plane-wave matrices appearing in the summation over reciprocal lattice vectors as 
\begin{eqnarray}
     W^{\eta}_{\textbf{G}\textbf{G}'}(\textbf{k}) =\left[
     \begin{tabular}{cc}
      $w^{\eta, \uparrow\uparrow}_{\textbf{k} + \textbf{G}} $  &  $w^{\eta, \uparrow\downarrow}_{\textbf{k} + \textbf{G}} $ \\
       $w^{\eta, \downarrow\uparrow}_{\textbf{k} + \textbf{G}} $ &  $w^{\eta, \downarrow\downarrow}_{\textbf{k} + \textbf{G}}$
     \end{tabular}\right],   \label{eq4}
\end{eqnarray}
whose shape is to be determined from symmetry analysis, as described next.

\begin{figure}[t]
\includegraphics[width=\columnwidth]{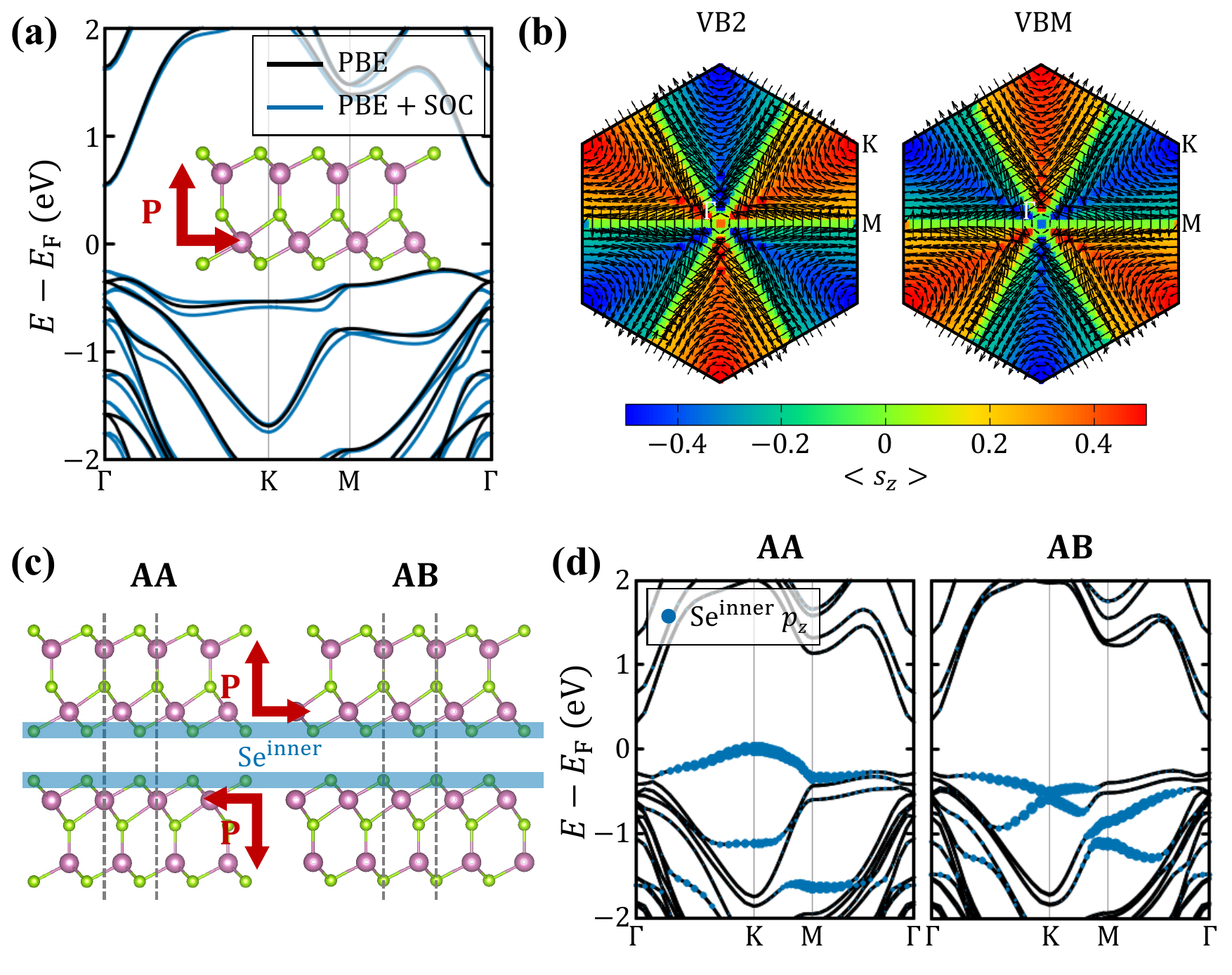}
\caption{(a) Electronic structure of monolayer $\alpha$-In$_2$Se$_3$ without
(black) and with (blue) spin-orbit coupling (SOC). Its crystal structure and polarization vector was overlaid. (b) The momentum-resolved spin textures computed from the
highest (VBM, right) and second highest (VB2, left) valence bands. The size of black arrows and color coordinates indicate in-plane and out-of-plane spin components, respectively. (c)
Side views of AA and AB stacking configurations of bilayer  $\alpha$-In$_2$Se$_3$. Se atoms placed on the interlayer region (Seinner) were highlighted with blue color. Interlayer distances in the
two stacking configurations were fixed to be $3.3$ \AA. (d) Electronic structures of AA and AB stacking configurations with SOC. Radius of blue sphere represent the contribution from
$p_z$ orbitals of Se$^{\textrm{inner}}$ placed atoms}
\label{fig1}
\end{figure}

\begin{figure*}[t]
\includegraphics[width=2.0\columnwidth]{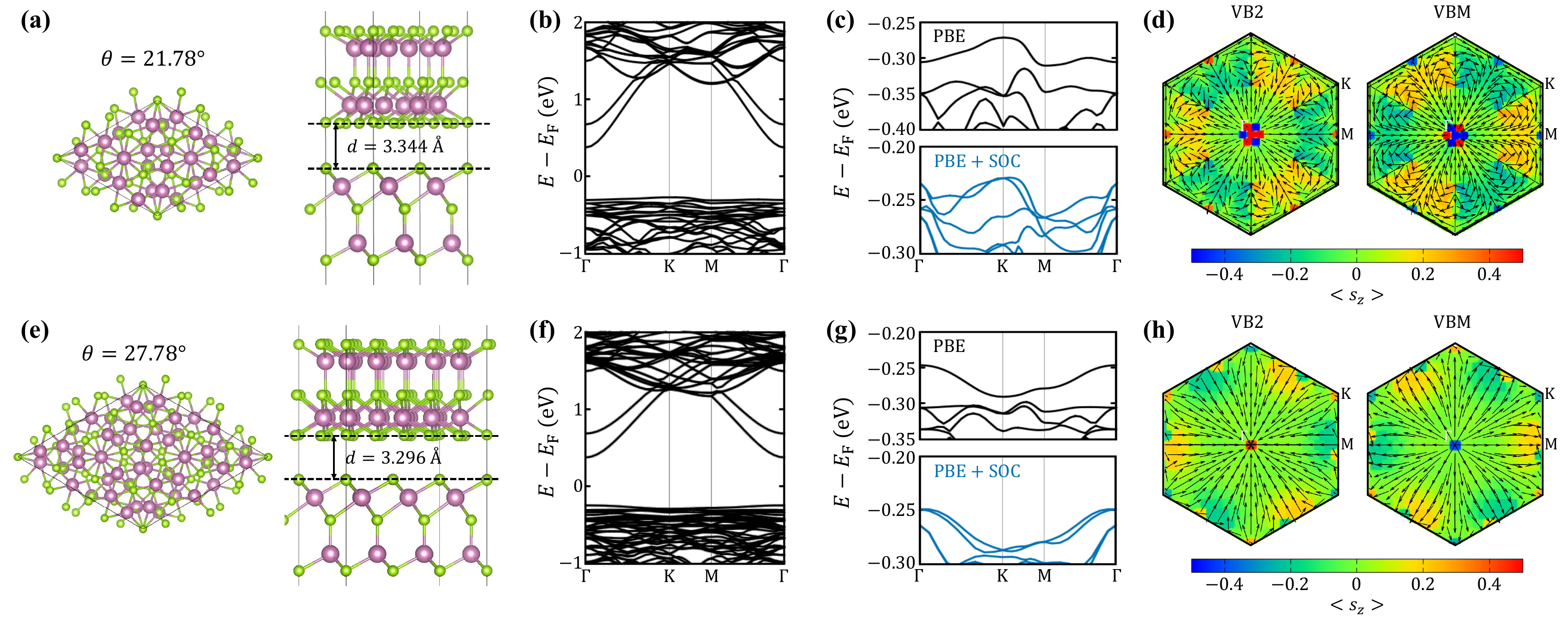}
\caption{(a) Top and side views of $\theta = 21.78^{\circ}$ twisted $\alpha$-In$_2$Se$_3$ bilayer. 
(b) Its corresponding electronic structure without SOC and (c) detailed electronic structures near valence band maximum without (upper, black) and with (lower, blue) SOC. 
(d) The momentum-resolved spin textures computed from the highest (VBM, right) and second highest (VB2, left) valence bands. The size of black arrows and color coordinates indicate in-plane and out-of-plane spin components, respectively.
(e-h) The top and side views, the electronic structures, and the momentum-resolved spin textures of $\theta = 27.78^{\circ}$ twisted $\alpha$-In$_2$Se$_3$ bilayer.
}
\label{fig3}
\end{figure*}

Although a relative twist between the two layers breaks all mirror symmetries due to the induced structural chirality, we note that a combination of the monolayer mirror symmetries and a \textit{horizontal mirror} operation, $\mathcal{M}_{xy}$, will leave the twisted moir\'e lattice invariant. Such observation is key in capturing the chirality of the twisted structure, which requires at least two successive mirror operations as a symmetry, e.g., while the first mirror operation will reverse the handedness or chirality of the twisted bilayer, applying a second mirror operation will flip the handedness back to its original configuration. Such a composite symmetry operation dictates the form of the Moir\'e coupling due to structural chirality. To show this, we proceed by enforcing that the twisted bilayer Hamiltonian remains invariant under $g_x = \mathcal{M}_{xy}\mathcal{M}_{xz}$ and $g_y = \mathcal{M}_{xy}\mathcal{M}_{yz}$ next~\cite{Snote}, where we have considered a general 2D linear-in-\textbf{k} form $W^{\eta}(\textbf{k}) = w_0s_0 + \sum_{\alpha\beta = x,y} w^{\eta}_{\alpha\beta}k_{\alpha}s_{\beta}$. Noticing that $g_x \eta = \eta$ and $g_y \eta = -\eta$ we obtain: $ s_x W^{\eta,\dagger}_{g_x\textbf{G}, g_x\textbf{G}'}(g_x\textbf{k}) s_x = W^{\eta,\dagger}_{\textbf{G} \textbf{G}'}(\textbf{k})$ and $ s_y W^{-\eta,\dagger}_{g_y\textbf{G}, g_y\textbf{G}'}(g_y\textbf{k}) s_y = W^{\eta,\dagger}_{\textbf{G} \textbf{G}'}(\textbf{k})$. Enforcing these relations, we find that the $g_x$ symmetry requires that $w^{\eta}_{xy} = w^{\eta}_{yx} = 0$. Further, $C_{3z}$ symmetry ensures $w^{\eta}_{xx} = w^{\eta}_{yy}$, while $g_y$ symmetry requires that $w^{\eta}_{xx} = w^{-\eta}_{xx}$. Therefore, 
\begin{eqnarray}
    W^{\eta}_{\textbf{G}\textbf{G}'}  = w_0 s_0 + w^{\eta}_{xx} (\textbf{k} + \textbf{G})\cdot\textbf{s}, \label{eq66}
\end{eqnarray}
 with $W^{\eta}_{\textbf{G}\textbf{G}'} = W^{-\eta}_{\textbf{G}\textbf{G}'}$, is the symmetry allowed shape of the interlayer moir\'e coupling to linear order in $\textbf{k}$, where $\textbf{s} = (s_x, s_y)$ is the vector of spin Pauli matrices. Equation~(\ref{eq66}) is one of the central results of this paper. It summarizes the impact of structural chirality on the electronic states through a \textit{Weyl-type interlayer moir\'e coupling in strongly spin-orbit coupled twisted bilayers}. Further analysis reveal that a general $C_{nz}$ symmetry constrains the coefficients of a $g_x$ and $g_y$ symmetric interlayer moir\'e coupling of the form $w_{xx} k_x s_x + w_{yy} k_y s_y$ to obey $(w_{xx} - w_{yy})\sin(2\pi/n) = 0$~\cite{Snote}. Hence, $C_{3z}$, $C_{4z}$ and $C_{6z}$ symmetric moir\'e patterns will enforce $w_{xx} = w_{yy}$, giving rise to Eq.~(\ref{eq66}). In $C_{2z}$ symmetric twisted bilayers, there are no extra constrains over $w_{xx}$ and $w_{yy}$. Therefore, the system can also support chiral radial-tangential~\cite{PhysRevB.104.104408} interlayer moir\'e couplings. Figure~\ref{fig2}(c) schematically illustrates this situation; The ideal radial (top panel) and the radial-tangential (middle panel) spin textures are the only possible spin textures under $g_x$ and $g_y$ symmetry. Here, the first (vertical) mirror operation $\mathcal{M}_{yz}$ flips all spin states at each momentum, which is recovered back after a second (horizontal) mirror operation, i.e., it respects $g_y$. In contrast, the pure Rashba spin texture (bottom panel) is not left invariant under the $g_y$ and, therefore, the interlayer moir\'e coupling cannot support Rashba-like spin texture.

 Finally, the moir\'e Hamiltonian can be explicitly written as
 \begin{eqnarray}
     H^{\eta}_{\textbf{G}\textbf{G}'}(\textbf{k}) = \gamma |\textbf{k} - \textbf{G}|^2 \delta_{\textbf{G}\textbf{G}'} + \nonumber \\ \sum_{j=1,2,3}W^{\eta}_{j}(\textbf{k})(\delta_{\textbf{G}-\textbf{G}',\textbf{q}_j} + \delta_{\textbf{G}-\textbf{G}',-\textbf{q}_j}),
     \label{eq55}
\end{eqnarray}
in the basis $u^{\eta}_{n\textbf{k}} = \sum_{l} u^{\eta l}_{n\textbf{k}}$, where $l = T,B$ is the layer index. It is noteworthy that any contributions to a Rashba texture originating from the bare electron states within each monolayer cancels out exactly, as they are equal and opposite due to the assumed inversion symmetry. The electronic structures without ($w^{\eta}_{xx} = 0$) and with ($w^{\eta}_{xx} \neq 0$) spin-orbit coupling are shown in Fig.~\ref{fig2}(d). The presence of spin-orbit interaction lifts the band degeneracies everywhere in momentum space, except at TRIM points in the moir\'e Brilluion zone, giving rise to semimetallic band crossings. The TRIM degeneracies are ensured by time-reversal symmetry through Kramers theorem, in full analogy with the three-dimensional topological chiral crystals hosting Kramers-Weyl fermions~\cite{Chang2018}, but now in 2D. Further, these 2D moir\'e Kramers-Weyl fermions exhibit ideal radial spin texture, originating from the Weyl-type interlayer moir\'e coupling. Figure~\ref{fig2}(e) displays the momentum resolved spin texture of the first two valence states, dubbed VBM and VB2 in Fig.~\ref{fig2}(d). As expected, the states around the time-reversal induced semimetallic band crossings at TRIM points, such as K, K' and $\Gamma$ points, exhibit ideal radial spin texture. Equation~(\ref{eq66}) also indicates that the structural chirality of the twisted bilayers guarantees purely radial 2D spin texture at \textit{any twist angle}\cite{note}. Next, we present our first principles calculations showing that the physics discussed above is realized in $\alpha$-In$_2$Se$_3$ bilayers.

\emph{Crystal and electronic structures of mono and bilayer $\alpha$-In$_2$Se$_3$. } 
Figure~\ref{fig1}(a) shows the crystal and electronic structures of monolayer $\alpha$-In$_2$Se$_3$ calculated by first-principles calculations based on the density functional theory (DFT)~\cite{Kohn1965,Kresse1996} (See Note~\ref{note1} in Supplementary Information for details).
The monolayer $\alpha$-In$_2$Se$_3$ belongs to the $P3m1$ polar space group (no. 156)~\cite{ding2017prediction}. 
Its quintuple atomic layer structure can be classified into octahedra (Oct, InSe$_6$) and tetrahedra (InSe$_4$) sites,~\cite{tao2022designing} and thus it exhibits both in-plane and out-of-plane ferroelectricity (Inset in Fig.~\ref{fig1}) which survives at room temperature~\cite{cui2018intercorrelated}. 
Within PBE XC functional, in the absence of spin-orbit coupling (SOC), the electronic spectrum of such systems has an indirect bandgap of 0.79~eV. In addition, while its conduction band has a sharp parabolic shape, its valence bands are rather flat and are mainly contributed from $p$ orbital of Se at the Oct site.
Due to the broken inversion symmetry, even without SOC, $\alpha$-In$_2$Se$_3$ exhibits a strong orbital Rashba effects~\cite{park2011orbital,lee2020unveiling} mainly in valence bands due to its orbital characters.~\cite{Snote}
When spin-orbit coupling (SOC) is turned on, the predominant orbital Rashba texture evolve into a spin Rashba texture, resulting in a sizable spin splitting in the valence bands.
To clarify this, we displayed the momentum-resolved spin textures of the highest and second highest valence bands (VBM and VB2) in Fig~\ref{fig1}(b). Our analysis across the entire Brillouin zone revealed helical spin textures with opposing chiralities in the two bands, a distinctive signature of Rashba spin splitting. Note that the Rashba effects is rather weak in the conduction band because its orbital character cannot produce the orbital Rashba effects. Therefore, in the following paragraphs, we mainly focus on the valence bands of this system.

We discuss now the bilayer $\alpha$-In$_2$Se$_3$. Figure~\ref{fig1}(c) shows the side views of two distinct stacking configurations of bilayer $\alpha$-In$_2$Se$_3$, referred to as ``AA'' and ``AB''.
Note that we only consider the bilayer with two Oct sites facing each other due to its lower stacking energy compared to the other stacking configurations~\cite{cui2018intercorrelated, PhysRevLett.126.057601}.
In both configurations, the non-zero dipole moments originating from each layer were canceled out, restoring inversion symmetry.
Without any constraint, ``AA'' stacking has much wider interlayer distance (3.761~{\AA}) than ``AB'' stacking (2.895~{\AA}).~\cite{Snote}
However, both ``AA'' and ``AB'' regions with an intermediate interlayer distance between two configurations locally emerge in the twisted bilayer.~\cite{tao2022designing} 
Therefore, here, we fixed an interlayer distance to 3.3~{\AA} and evaluated their electronic structures as shown in Fig.~\ref{fig1}(d).
In monolayer $\alpha$-In$_2$Se$_3$, the low energy valence bands originate from the $p_x$, $p_y$ and $p_z$ states of the Se atoms pertaining to the Oct sector of the unit cell, which we refer to as ``\textit{inner}" in Fig.~\ref{fig1}(c). Furthermore, at the K and K' points ($\Gamma$ point), valence states have predominantly $p_z$ ($p_x \pm i p_y$) character, henceforth referred to as $p_z$Se-Oct states ($p_{x,y}$Se-Oct). Our first principles calculations reveal that, while $p_z$Se-Oct states in an AB-stacked $\alpha$-In$_2$Se$_3$ bilayer form a linear band crossing at the K points, the AA-stacked configuration supports a strong $p_z$Se-Oct level repulsion at K (and K'), causing the appearance of valleys at the corners of the Brillouin zone whose low energy physics is dominated by $p_z$Se-Oct anti-bonding states from the two layers~\cite{Snote, tao2022designing}. This is clearly shown in Fig.~\ref{fig1}(d). Therefore, the low energy description of moir\'e physics in twisted $\alpha$-In$_2$Se$_3$ bilayers must be constructed over these strongly spin-orbit coupled $p_z$ states, connecting to the continuum model presented in the previous section.

\emph{First Principles Calculations of Twisted $\alpha$-In$_2$Se$_3$ Bilayers.}
We performed fully relativistic DFT calculations in twisted $\alpha$-In$_2$Se$_3$ bilayers with distinct twist angles. Figure~\ref{fig3}(a) shows the relaxed crystal structure of  $\theta = 21.78^{\circ}$ twisted $\alpha$-In$_2$Se$_3$ bilayer. The interlayer distance was evaluated to be 3.344~{\AA} which agrees with our prediction in the previous section. Similar with the untwisted case, as shown in Fig.~\ref{fig3}(b), its conduction bands exhibit a parabolic shape while the valence bands are flat. Then, we visualized detailed structures of the valence bands with and without SOC, as shown in Fig.~\ref{fig3}(c).
As reported in an earlier study,~\cite{tao2022designing} we also found an isolated quasi-flat band at the valence band maximum in the case of without SOC. When the SOC is turned on, the spin degeneracy of the bands is lifted due to the structural chiralty, except at the TRIM.
In other words, the TRIM points host Kramers-Weyl nodes protected by time-reversal symmetry~\cite{Chang2018}.
Further, we indeed confirmed the perfect radial spin textures in the VBM and VB2, as visualized in Fig.~\ref{fig3}(d) and (h), consistent with our predictions from the continuum model. Similar band structures and spin textures, obtained from our first principles DFT results shown in Fig.~\ref{fig3}(e-h) and in Fig.~\ref{figs2}, are also found at other twist angles, indicating that these general properties are intrinsic features of electron states in these systems. This suggests that the moir\'e Kramers-Weyl physics presented here is universal, in full agreement with our continuum model prediction. 

\emph{Discussion and Conclusion.}
It is worth noting that the emergence of the radial spin texture in the twisted homobilayer is a generic phenomenon, as explicitly explained through our continuum model. More generally, other SOC contributions, such as valley-Zeeman or Kane-Mele SOC may contribute to Eq.~(\ref{eq1}) in addition to Rashba-type SOC, resulting in additional out-of-plane spin components, which has been recently predicted in twisted WSe$_2$ homobilayers~\cite{frank2024emergence}. 
In this case, competing Rashba and valley-Zeeman SOC determines the dominant spin texture (radial in-plane versus valley-dependent out-of-plane).
In this work, the intrinsic ferroelectric order in monolayer $\alpha$-In$_2$Se$_3$ strongly enhances Rashba SOC, resulting in the dominance of radial in-plane spin textures.
Moreover, our DFT calculations clearly show that additional interaction with other valence states, with $p_x\pm i p_y$ character, leads to the non-zero out-of-plane spin textures (See Fig.~\ref{fig3}(d, h) and Fig.~\ref{figs2}(d)). Therefore, the isolated flatband observed in the absence of SOC is also beneficial to suppress out-of-plane spin components.
The above discussion suggests that the twisted $\alpha$-In$_2$Se$_3$ bilayer is an optimal system for realizing ideal moir\'e Kramers-Weyl nodes with perfect radial spin textures.

In summary, we showed that two-dimensional Kramers-Weyl fermions are an emergent manifestation of structural chirality in strongly spin-orbit coupled twisted bilayers. The chirality inherent to the moir\'e pattern has been shown to give rise to a Weyl-type interlayer moir\'e coupling, from which the ideal radial spin texture of the moir\'e Kramers-Weyl fermions manifests at finite twist angle. Our findings are fully supported by fully relativistic first principles electronic structure calculations of a few twisted $\alpha$-In$_2$Se$_3$ bilayers systems, where the semimetallic band crossings as well as ideal radial spin textures are explicitly shown.

\textit{Acknowledgments.} 
S. L. is primarily supported by Basic Science Research Program through the National Research Foundation of Korea funded by the Ministry of Education (NRF-2021R1A6A3A14038837). S. L. and T. L. are partially supported by NSF DMREF-1921629. D. S. and T. L. acknowledge partial support  from Office of Naval Research MURI grant N00014-23-1-2567.

\bibliographystyle{apsrev}
\bibliography{my.bib} 

\begin{thebibliography}{53}
\bibitem[{not()}]{note}
\bibinfo{note}{We note that radial tangential spin-textures are also possible away from the $\Gamma$ and $K$ points in this system. This is due to the possibility of distinct wave vector point groups associated other high symmetry points, such as the $M$ point. See Ref.~\cite{PhysRevB.104.104408} for more details.}
\end{thebibliography}

\end{document}